# Segregation, integration and balance of large-scale resting brain networks configure different cognitive abilities

Rong Wang[1,2,3†], Mianxin Liu[2†], Xinhong Cheng[1], Ying Wu[3,4,5], Andrea Hildebrandt[6,7*] & Changsong Zhou [2,8*]

1. College of Science, Xi'an University of Science and Technology, Xi'an 710054, China.

2. Department of Physics, Centre for Nonlinear Studies, Institute of Computational and Theoretical Studies, Hong Kong Baptist University, Hong Kong.

3. School of Aerospace Engineering, Xi'an Jiaotong University, Xi'an, 710049, China.

4. State Key Laboratory for Strength and Vibration of Mechanical Structures, Xi'an Jiaotong University, Xi'an, 710049, China.

5. National Demonstration Center for Experimental Mechanics Education, Xi'an Jiaotong University, Xi'an, 710049, China.

6. Department of Psychology, Carl von Ossietzky Universität Oldenburg, 26129, Oldenburg, Germany.

7. Research Center Neurosensory Science, Carl von Ossietzky Universität Oldenburg, 26129, Oldenburg, Germany.

8. Department of Physics, Zhejiang University, Hangzhou, 310027, China.

[†]These authors contributed equally to this work

[*]Corresponding authors. Email: andrea.hildebrandt@uni-oldenburg.de (A.H.); cszhou@hkbu.edu.hk (C.Z.)

**Abstract**

Diverse cognitive processes set different demands on locally segregated and globally integrated brain activity. However, it remains unclear how resting brains configure their functional organization to balance the demands on network segregation and integration to best serve cognition. Here, we use an eigenmode-based approach to identify hierarchical modules in functional brain networks, and quantify the functional balance between network segregation and integration. In a large sample of healthy young adults (n=991), we combine the whole-brain resting state functional magnetic resonance imaging (fMRI) data with a mean-filed model on the structural network derived from diffusion tensor imaging and demonstrate that resting brain networks are on average close to a balanced state. This state allows for a balanced time dwelling at segregated and integrated configurations, and highly flexible switching between them. Furthermore, we employ structural equation modelling to estimate general and domain-specific cognitive phenotypes from nine tasks, and demonstrate that network segregation, integration and their balance in resting brains predict individual differences in diverse cognitive phenotypes. More specifically, stronger integration is associated with better general cognitive ability, stronger segregation fosters crystallized intelligence and processing speed, and individual's tendency towards balance supports better memory. Our findings provide a comprehensive and deep understanding of the brain's functioning principles in supporting diverse functional demands and cognitive abilities, and advance modern network neuroscience theories of human cognition.

**Introduction**

The brain dynamically reconfigures its functional organization to support diverse cognitive task performances[1,2]. Successful reconfiguration underlying better task performance relies not only on sufficiently independent processing in specialized subsystems (i.e., segregation), but also on effective global cooperation between different subsystems (i.e., integration)[1-5]. It has been observed that diverse cognitive tasks set different demands on segregation and integration[3,5-11]. Higher segregation has been linked to simple motor tasks, and higher integration seems to underlie performance on tasks with a heavy cognitive load[7-11]. However, it remains a great challenge to understand how the brain's functional organization is configured to support heterogeneous demands on segregation and integration for diverse cognitive processes.

Independently from specific task demands, the brain's functional organization at rest can mirror relevant task-induced activity patterns and thus predict task performance[12,13]. Emerging evidence suggests that smaller differences between functional patterns at rest versus task states can facilitate better cognitive performance[12-15]. Since diverse cognitive tasks differently demand on segregation and integration[3,5-11], the brain's functional organization at rest is expected to possess the intrinsic capability of supporting diverse cognitive processes. Furthermore, previous studies suggest that healthy resting brains operate near a critical state to render the capability of rapidly exploring and switching in the brain's state space with large operating repertoires[3,12,16-18]. Resting brains are thus supposed to balance the segregation and integration[16,19], so as to satisfy competing cognitive demands. However, this theory still lacks empirical evidence on whether large-scale brain networks at rest entail a balance between segregation and integration, and whether the functional balance is associated with individual differences in cognitive abilities.

To date, most of relationships between brain functional configurations at rest and cognitive abilities are based on single tasks[6-11] that assess only specific aspects of cognition. General and domain-specific cognitive abilities are modeled at the latent level based on multiple tasks in differential psychology[20-23]. These latent cognitive abilities generalize across tasks of the same domain and account for measurement error[24], and thus are much more suitable to reveal the neural basis of individual differences in human cognition. Recently, a cross-disciplinary Network Neuroscience Theory (NNT) proposed a general framework to investigate the neural basis of cognitive abilities relying on system-wide topology characteristics and dynamics of brain networks[25]. According to NNT, brain networks functioning in an "easy-to-reach state" serve crystallized intelligence, whereas a "difficult-to-reach state" is needed for fluid intelligence[25]. General cognitive ability, which is a statistical summary of fluid and crystallized intelligence, is considered to be facilitated by the capacity to flexibly switch between the above mentioned network states, i.e., an optimal balance between local and global processing[25]. These predictions are still built upon a fragile empirical basis[26]. Elucidating the relationship between functional balance and different cognitive abilities is crucial for validating and reframing NNT, especially with respect to the question of whether a functional balance in the brain at resting state benefits the general cognitive ability across individuals.

Before these questions can be robustly answered, the balance between segregation and integration in the large-scale brain must be explicitly defined and quantified. The modular structure of brain's functional connectivity (FC) networks is known to provide the basis for specialized information processing within modules and the integration between them[7,9,14,25,27,28]. Although many studies have applied measures based on modules at a single level to study functional segregation and integration[6,7,9,14,27], empirical evidence for a balance in the brain's functional organization is still lacking. In fact, brain FC networks are hierarchically organized[29-31]. Such an organization potentially supports nested segregation and integration across multiple levels. However, the classically applied modular partition at a single level does not allow the detection of hierarchical modules across multiple levels[32,33]. This insufficiency seems to be the main reason for the lack of a robust quantitative definition of the balance between segregation and integration.

Here, we explicitly identified the functional balance based on hierarchical modules of resting brain FC networks and explored associations with diverse cognitive abilities in a sample of 991 heathy young adults from WU-Minn Human Connectome Project (HCP)[34]. Using an our previously published method called nested-spectral partition (NSP) based on eigen-modes[18], we first detected hierarchical modules in FC networks to propose an explicit balance measure. Second, we combined real data and a Gaussian linear model to demonstrate the functional balance in the group-averaged brain at resting state. Then, we investigated individual differences in the balance and relationships to the temporal switching between segregated and integrated states. Finally, we applied structural equation modeling (SEM) to estimate latent factors of general and domain-specific cognitive abilities and investigated how segregation, integration and their balance configure them.

**Results**

**Hierarchical modules in FC networks.** Since the length of functional magnetic resonance imaging (fMRI) series affects the dynamic properties of FC networks[28], we concatenated the fMRI data across four scanning sessions and all individuals to obtain a stable average FC network consisting of $N = 360$ regions (Fig. 1a). The NSP method was applied to detect hierarchical modules in the FC network according to the functional modes (i.e., eigenvalues $\Lambda$ and eigenvectors $U$) which were sorted in descending order of $\Lambda$ (see Methods). At the first level, corresponding to the 1$^{st}$ eigenvector with the same sign for each region (Fig. 1b), the stable average FC network has the largest coactivation mode, effectively involving the whole brain in a single module (Fig. 1c). At the second level, brain regions are partitioned into two large modules that correspond to positive and negative signs in the 2$^{nd}$ eigenvector (Fig. 1b). This functional partition pattern nearly coincides with the division between anterior and posterior brain regions (Fig. 1c), suggesting that the 2$^{nd}$ mode reflects the modular division of the brain into anterior and posterior functional systems. Furthermore, according to the negativity and positivity of the 3$^{rd}$ eigenvector, each module at the second level was further subdivided into two modules at the third level (Fig. 1b). Successively, with the increasing order of functional modes, the FC network is modularly partitioned into multiple levels. The hierarchically partitioned FC network has higher average link weights within modules than between modules across multiple levels (Fig. 1a), clearly manifesting hierarchical modules of the FC network.

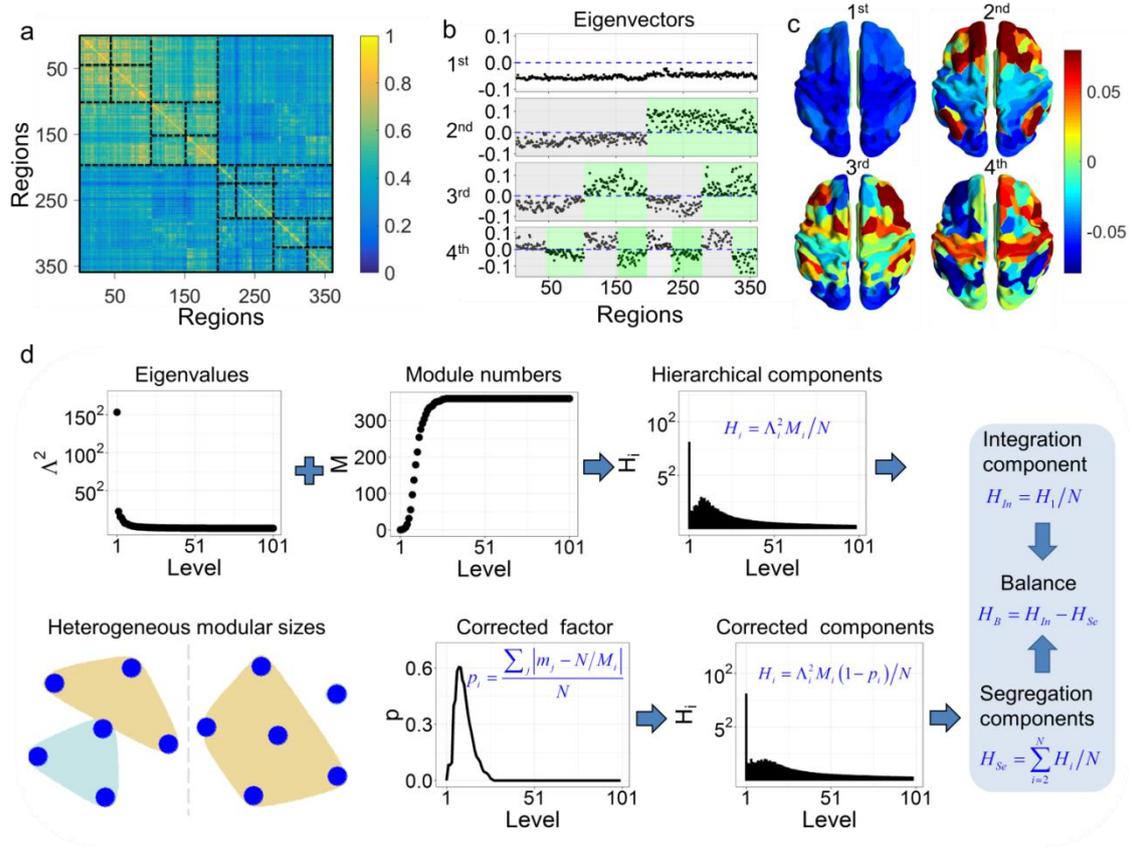

**Fig. 1. Hierarchical segregation and integration in FC networks.** (a) The stable average FC network and its hierarchical modules. Black dashed lines represent the boundaries of hierarchical modules suggested by the NSP method. (b) Hierarchical modular partition in the first four functional modes, where modules in each level (gray and green) are detected according to the positivity or negativity of eigenvector components. (c) Spatial patterns corresponding to the first four functional modes. (d) The pipeline of defining hierarchical segregation and integration (see Methods). The combination of contribution $\Lambda^2$ and module number $M$ provides the hierarchical functional components $H_i$ at different levels. Since the modular size may be heterogeneous even for the same module number in each level (see an example of six nodes partitioned into two modules, where the modular sizes of 1 and 5 generate higher integration and lower segregation than the sizes of 3 and 3), the hierarchical components $H_i$ need to be corrected. Then, the first level contributes to the global integration, and the $2^{nd}$-$360^{th}$ levels contain the multilevel segregation components. A functional balance is defined when the global integration component equals the total segregation component.

**Hierarchical segregation and integration in FC networks.** Hierarchical modules of brain FC networks involve hierarchically segregated and integrated interactions between regions. At a specific level, regions with the same sign of eigenvector components (e.g., negative or positive) within a module are jointly activated to achieve functional integration, whereas the opposite activation (e.g., negative and positive signs) of regions indicates segregation between modules. The integration of smaller segregated modules at a high-order level (e.g., $i$th mode) leads to a formation of larger module at the lower-order level (i.e., $(i-1)$th mode), which further generates segregation with other large modules at this lower-order level. Thus, functional segregation and integration are intricately interrelated and hierarchically organized in a nested manner across multiple levels. We defined a weighted module number $H_i$ (Eq. (2)) to quantify the nested segregation and integration (see Methods). The functional component at the first level measures the degree of global integration and is denoted as the *integration component* $H_{In}$ (Eq. (4)). Functional components across all higher levels (i.e., $2^{nd}$-$360^{th}$ levels) quantify the hierarchical segregation and are summed to obtain an overall measure of the *segregation component* $H_{Se}$ (Eq. (5)).

Aiming to confirm the validity of $H_{In}$ and $H_{Se}$, we implemented a Gaussian linear process on structural connectivity (SC) networks and produced simulated individual FC matrices for sufficiently long time frames[18,35] (see Methods). At an intermediate coupling in the model (i.e., $c=70$), the simulated FC networks were most similar to the empirical FC network. This similarity is indicated by the same mean correlation, a minimal distance between real and simulated FC matrices, and the minimal difference of regional degrees, the same characteristic path length, clustering coefficient and global efficiency (Figs. 2c, d and Supplementary Fig. 2). These results suggest that resting brains correspond to the dynamic point at the critical coupling ($c=70$) in the Gaussian model.

For small couplings (e.g., $c=20$), brain regions are relatively independent and form sparse FC networks (Fig. 2a), as indicated by correlation values approaching zero (Fig. 2b). This state is only able to support segregated activity and is not sufficient for large-scale integration. Correspondingly, segregation components $H_i(i \geq 2)$ have high values, while the global integration component is small (Fig. 2b). In contrast, for strong couplings (e.g., $c=120$), brain regions are strongly and densely connected to form globally synchronized patterns (Fig. 2a), as indicated by correlation values distributed towards one (Fig. 2b). This large-scale synchronization recruits the whole brain, exhibiting a high integration component $H_1$ and small segregation components (Fig. 2b). This state does not allow specialized activity. Thus, during the dynamic transition from asynchronous to synchronous states, global integration increases and segregation decreases, which is consistent with classical graph-based measures of decreased modularity and increased participation coefficient (Figs. 2e, f). Crucially, this dynamic transition can be well described by an increased $H_{In}$ and a decreased $H_{Se}$ (Fig. 2g), indicating the effectiveness of $H_{In}$ and $H_{Se}$ in characterizing segregated and integrated activity, respectively.

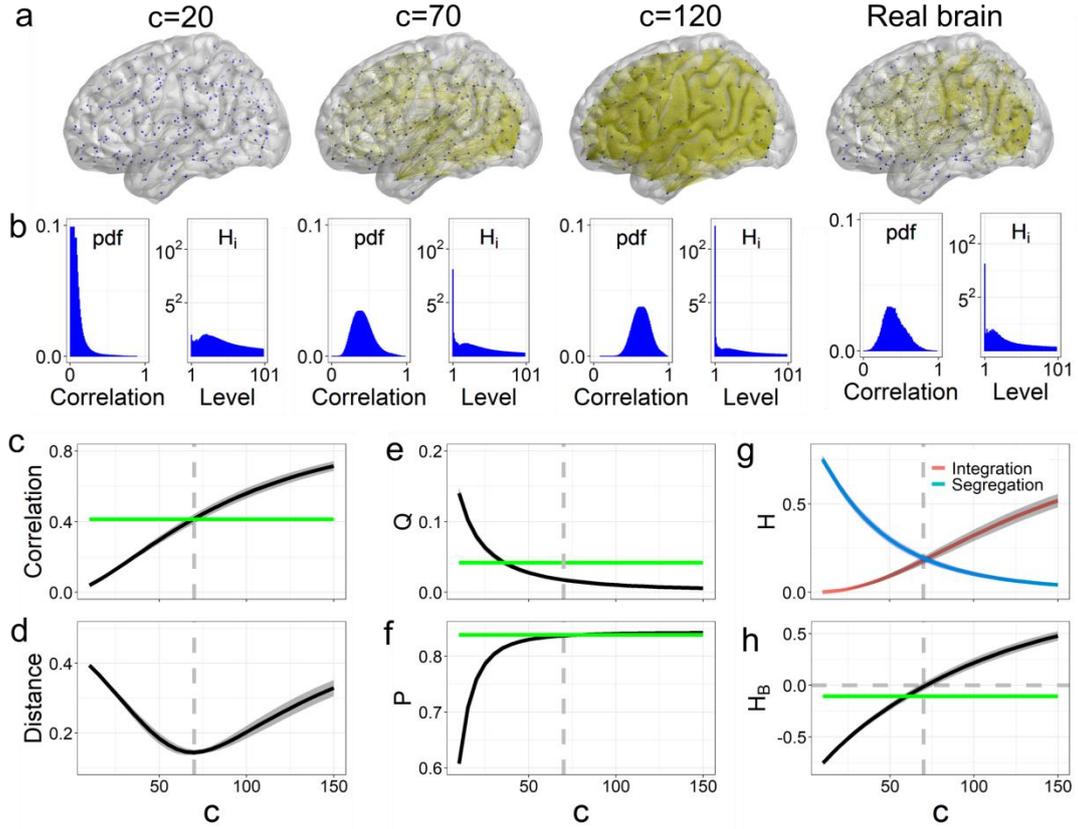

**Fig. 2. Balance between hierarchical segregation and integration. (a)** Simulated and real stable FC networks visualized using BrainNet Viewer[36]. The binarizing threshold was 0.65. **(b)** The probability density function (pdf) of correlation values in brain FC matrices and the hierarchical components $H_i$ in each level. **(c, d)** Mean correlation coefficients of simulated FC matrices and the Euclidean distance between the real stable FC matrix and simulated individual FC matrices at different $c$. **(e, f)** The modularity $Q$ and participation coefficient $P$ based on the seven functional subsystems (Supplementary Fig. 3) at a single level[37]. **(g, h)** The integration component $H_{In}$, segregation component $H_{Se}$, and balance predictor $H_B$ vary with $c$. Here, the shadows indicate the standard deviation across individuals, the horizontal green lines represent the corresponding values in the real stable FC network, and the vertical dashed lines mark the critical coupling ($c = 70$).

**Segregation-integration balance in large-scale resting brains.** Interestingly, the curves of $H_{In}$ and $H_{Se}$ in the Gaussian model intersect at the critical coupling $c = 70$ (Fig. 2g). Thus, the competition between integration and segregation $H_B = H_{In} - H_{Se}$ increases from negative values to positive values, and crosses zero at $c=70$ (Fig. 2h), indicating a theoretical balance between segregation and integration in the model. This balanced state is not revealed by the monotonically changed modularity and participation coefficient based on single-level modules (Figs. 2e, f). Most importantly, resting brains of heathy young adults are close to the balanced state with $H_B$ in the real stable FC network approaching zero ($H_B = -0.106$, see Fig. 2h). Indeed, fMRI signals inevitably contain measurement noise originating from various sources other than neural activity, which would artificially bring more segregation components into the real FC network. However, resting brains correspond to the critical coupling ($c=70$) in the model wherein the balance between segregation and integration theoretically exists. Thus, our results provide theoretical and empirical evidence that, at the population level, healthy young brains at rest tend to maintain a balance between segregation and integration.

**Individual differences in segregation-integration balance.** To study individual differences in segregation-integration balance, we constructed individual static FC networks from four concatenated fMRI sessions. The segregation and integration components in individual FC networks were calibrated to overcome the effects of a shorter fMRI series on segregation, integration and their balance (see Methods and Supplementary Fig. 4). This calibration restores the balance at the between person level and is appropriate to investigate the intrinsic relationship between brain measures and cognitive abilities (see Supplementary Table 1).

In individuals with sparse FC networks, brain regions are relatively separated with respect to their functional activation, and thus they generate a strong segregation component (large negative $H_B$, Figs. 3a-c). In contrast, brains with dense FC networks are highly integrated, corresponding to a strong integration component and large positive $H_B$. Put differently, individual brains with overly sparse or overly dense FC networks do not display a balance between segregation and integration.

However, brains with an intermediate density of FC networks are in a balanced state, with $H_B \approx 0$ (Figs. 3a-c). Across individuals, the modularity and participation coefficient relate to $H_B$ in a nonlinear and fuzzy manner (Figs. 3d, e and Supplementary Fig. 5). Importantly, these coefficients vary substantially for a specific $H_B$ value, particularly in segregated brains, indicating that $H_B$ based on hierarchical modules more precisely identifies the balanced state and individual differences therein than the single-level measures. Thus, $H_B$ may offer a more effective representation of an individual's tendency towards segregation vs. integration, with greater potential to be associated with cognitive abilities.

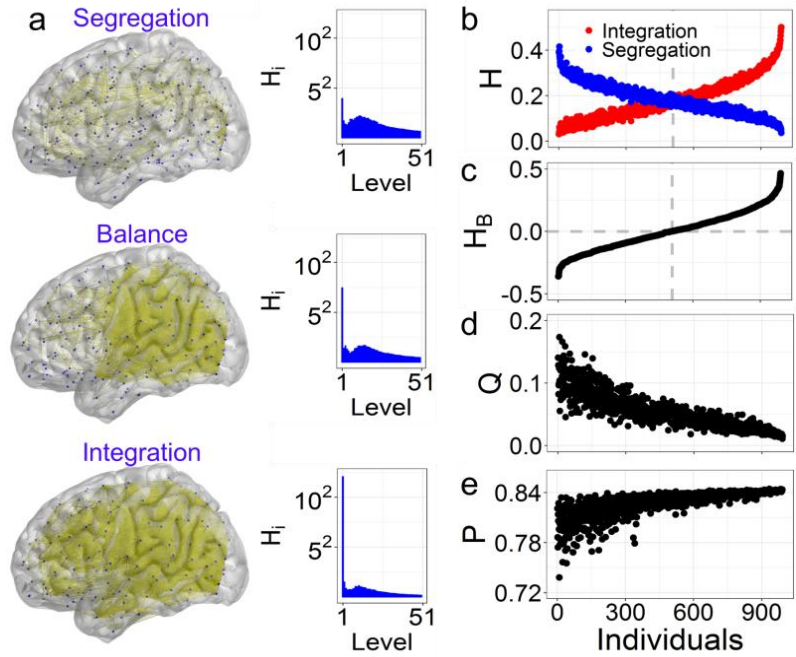

**Fig. 3. Individual differences in segregation-integration balance.** (**a**) Brain FC networks visualized for three individuals with a tendency towards segregation, balance and integration. The binarizing threshold was 0.65. The corresponding hierarchical components $H_i$ are also displayed. (**b, c**) The calibrated segregation component $H_{Se}$ and integration component $H_{In}$, as well as the balance indicator $H_B$ for all individuals who were sorted according to increasing values of $H_B$. (**d, e**) The corresponding individual modularity and participation coefficient.

**Balance supports flexible dynamic transition between segregated and integrated states.** To investigate the temporal switching between segregated and integrated states, we computed temporally dynamic FC networks. The mean segregation and integration components in dynamic FC networks for each individual were also calibrated to the corresponding individual static values (see Methods), such that the data-length-independent measure of flexible transition between different states were obtained.

The patterns of switching between segregated and integrated states differ significantly between individuals (Fig. 4a). For individual brains with static $H_B < 0$, most dynamic processes occur in the segregated state (i.e., $H_B(t) < 0$, see Fig. 4a), accompanied by a long dwell time $T_{Se}$ (Eq. (9), see Fig. 4b). In contrast, individual brains with static $H_B > 0$ have a long dwell time $T_{In}$ in the integrated state (i.e., $H_B(t) > 0$, see Figs. 4a, b). The static $H_B$ values across individuals were found to range from negative to positive values. As such, the brain exhibits a competition of dwell times between increased $T_{In}$ and decreased $T_{Se}$, as marked by $T_B = T_{In} - T_{Se}$ comprising a range of negative to positive values. Crucially, for individual brains with static $H_B \approx 0$, the dwell times in the integrated and segregated states are nearly equal, with a strong linear correlation between $T_B$ and $H_B$ across individuals ($r = 0.973$), where $T_B \approx 0$ matches $H_B \approx 0$ (Fig. 4c). These findings indicate the coexistence of a static and dynamic balance between segregated and integrated states.

Individual brains with high segregation or integration do not readily switch between segregated and integrated states (Fig. 4a). Contrarily, individual brains with static $H_B \approx 0$ exhibit apparently more frequent state transitions, as characterized by the highest switching frequency $f_{IS}$ (see Eq. (10)), and brains tending towards segregation or integration exhibit reduced $f_{IS}$ (see Fig. 4d). Thus, a balanced brain is most flexible in its dynamic transitions between segregated and integrated states.

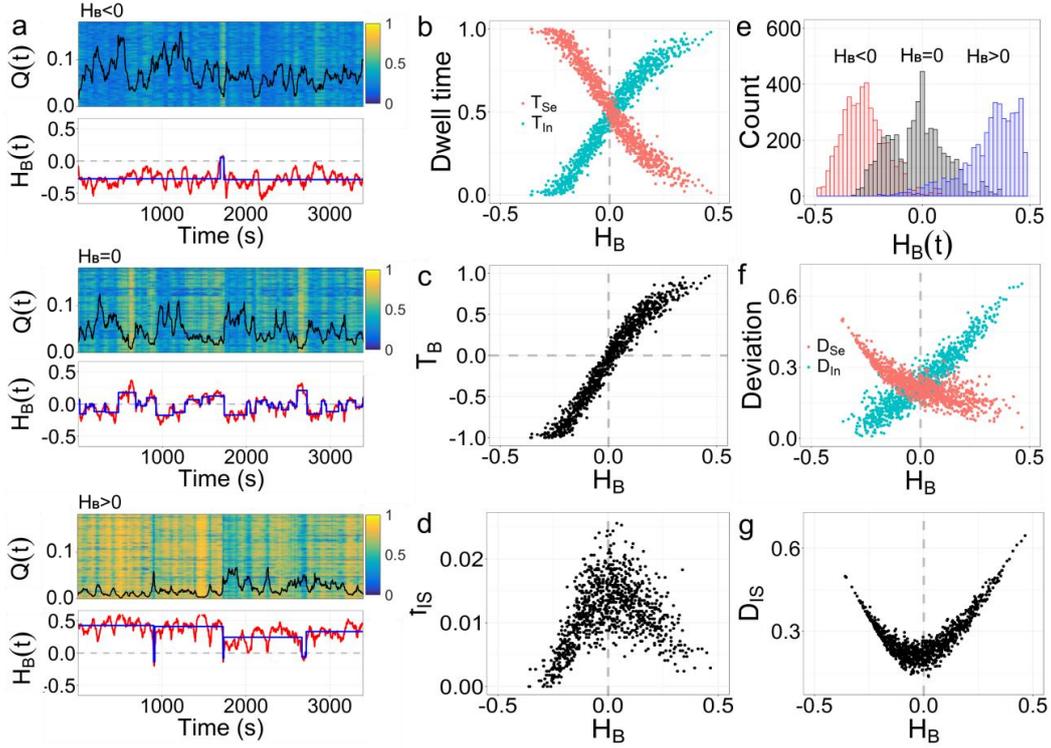

**Fig. 4. Functional balance supports more frequent state transitions. (a)** The activation patterns, temporal modularity $Q(t)$ (black line) and $H_B(t)$ (red line) of dynamic FC networks for three individuals (see Fig. 3a) tending towards segregation (upper, $H_B < 0$), balance (middle, $H_B = 0$) and integration (lower, $H_B > 0$). Here, the activation of a region is measured by the node degree. The blue lines mark the transition between segregated and integrated states. **(b, c)** The dwell time in the integrated state ($T_{In}$) and segregated state ($T_{Se}$), as well as the time difference $T_B$ which is positively correlated to $H_B$ across individuals. **(d)** The switching frequency $f_{IS}$ between segregated and integrated states. **(e)** The distributions of $H_B(t)$ for the three individuals. **(f, g)** The deviation degree from the balance to segregated state ($D_{Se}$) and integrated state ($D_{In}$), as well as the total deviation $D_{IS}$.

Furthermore, brains with higher segregation or integration substantially deviate from the balanced state during the switching process, whereas the deviation for the brain with static $H_B \approx 0$ is relatively small (Fig. 4a). With increasing static $H_B$, the distribution of $H_B(t)$ shifts from large negative to large positive values (Fig. 4e), and it is approximately zero for $H_B = 0$, reflecting the minimal deviation from the balanced state. To further confirm these results, we defined a degree of deviation from the balance to segregated or integrated states ($D_{Se}$ and $D_{In}$, see Eq. (11)). Individual brains with a strong tendency towards segregation (static $H_B < 0$) more strongly deviate towards a segregated state, with large $D_{Se}$. Brains with static $H_B > 0$ deviate towards integrated states, with large $D_{In}$ (Fig. 4f). For the brain with $H_B \approx 0$, $D_{In}$ and $D_{Se}$ are approximately equal (Fig. 4f), indicating equal deviation from balance towards segregated and integrated states. More importantly, the smallest total deviation $D_{IS} = D_{In} + D_{Se}$ is observed for brains with static $H_B \approx 0$, whereas the total deviation is increased for brains with a tendency towards segregation or integration (Fig. 4g). This deviation reflects a balanced competition between segregated and integrated states during dynamic reconfigurations to obtain an overall balanced brain.

**Segregation, integration and their balance predict different cognitive abilities.** To study how segregation, integration and their balance are associated with different cognitive abilities, we used SEM. We estimated latent factors of general and three domain-specific cognitive abilities from nine specific task performance indicators, spanning reasoning, crystallized intelligence, processing speed and memory (see Methods).

We first separately tested linear relationships between different network measures (i.e., $H_B$, $H_{Se}$ and $D_{Se}$, see Supplementary Table 1 for further associations) and cognitive ability factors in the entire sample (Fig. 5a and Supplementary Fig. 6). Three cognitive abilities (i.e., general cognitive ability, crystallized intelligence and processing speed) are significantly associated with the brain measures (Fig. 5b). First, the general cognitive ability factor is positively associated with $H_B$ (Standardized estimate coefficient $\beta = 0.087$, $p = 0.037$) and negatively associated with $H_{Se}$ ($\beta = -0.113$, $p = 0.007$) and $D_{Se}$ ($\beta = -0.155$, $p < 0.001$). Second, the crystallized intelligence factor is negatively related to $H_B$ ($\beta = -0.125$, $p = 0.016$) and positively related to $H_{Se}$ ($\beta = 0.148$, $p = 0.005$) and $D_{Se}$ ($\beta = 0.166$, $p = 0.002$). Third, the processing speed factor is negatively associated with $H_B$ ($\beta = -0.097$, $p = 0.016$) and positively associated with $H_{Se}$ ($\beta = 0.093$, $p = 0.022$) and $D_{Se}$ ($\beta = 0.089$, $p = 0.029$). Thus, a higher general cognitive ability relates to stronger integration, whereas greater segregation supports better crystallized intelligence and processing speed. Importantly, equivalent brain-behavior associations are not obtained using graph-based network measures at a single level (Supplementary Table 1). Our results emphasize the advantages of hierarchical module analysis for understanding the neural basis of individual differences in cognitive abilities.

Notably, the memory factor is not linearly associated with any of the considered brain measures (Fig. 5a and Supplementary Fig. 6), potentially implying a nonlinear relationship such that memory may be most strongly facilitated by the functional balance. To test this hypothesis, we partitioned the entire sample into groups of segregated (SG), balanced (BG) and integrated (IG) individuals, and investigated latent ability differences between them by means of multiple group SEM (see Methods). For a specific partition (Fig. 5c), the latent means of cognitive abilities in the three groups are significantly different ($p = 0.048$). In detail, the latent mean of general cognitive ability monotonically increases from the SG to IG. The latent differences between the SG and BG and between the BG and IG have small effect sizes (Cohen's $d = 0.25$ and 0.38), suggesting the highest general cognitive ability in the IG. In contrast, the latent mean of crystallized intelligence decreases from the SG to IG. The difference between the SG and BG reveals a medium effect size ($d = 0.47$) and the difference between the BG and IG is small ($d = 0.26$), indicating the better crystallized intelligence in the SG. The same trend of a monotonic decrease occurs for processing speed from the SG to IG. Latent differences between the SG and BG, as well as between the BG and IG, reveal small effect sizes ($d = 0.26$ and $0.21$), indicating the highest processing speed in the SG. These group differences are consistent with the linear associations estimated in the entire sample (Fig. 5b). Most importantly, the largest latent mean of memory was observed in the BG, whereas memory performance is smaller in both the SG and IG. The differences between the SG and BG, as well as between the BG and IG, indicate medium effects ($d = 0.46$ and $0.45$), supporting the highest memory in the BG.

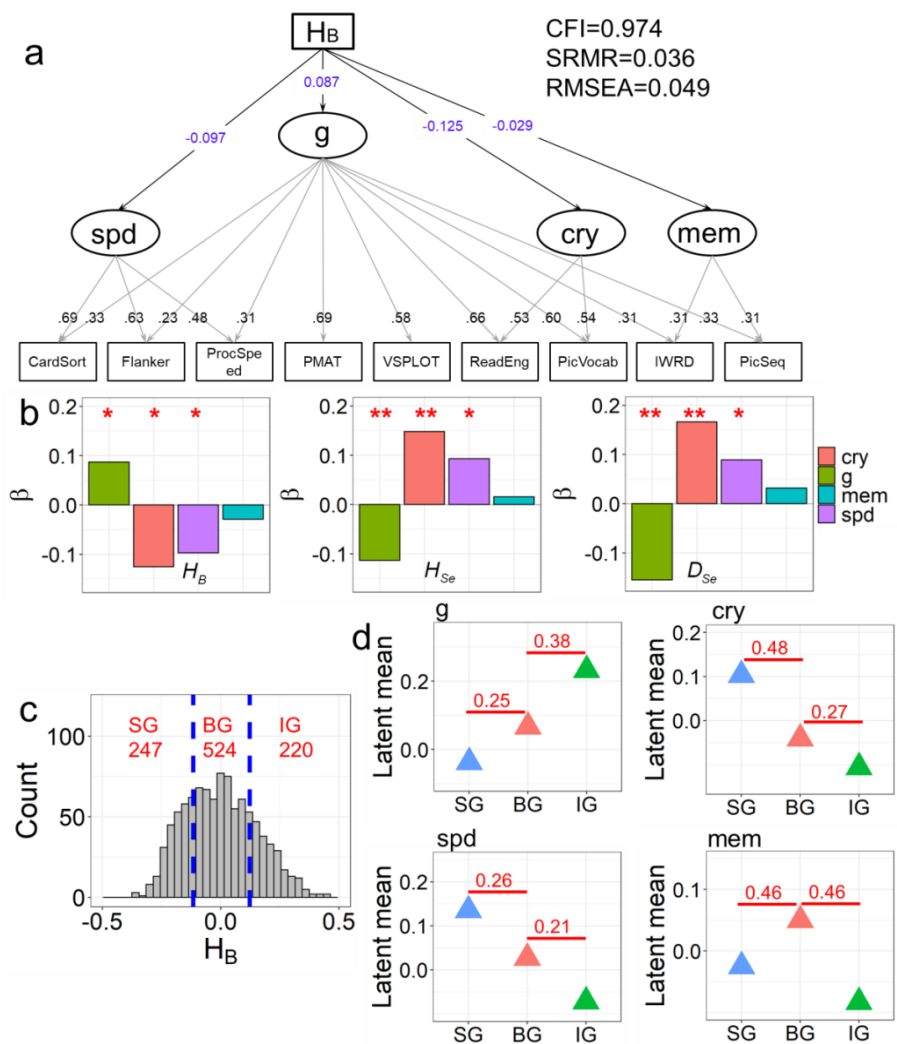

**Fig. 5. Brain-behavior relationship in SEMs. (a)** Schematic representation of the SEM testing the relationship between $H_B$ and cognitive abilities. Here, nine cognitive performance indicators (see Methods) were included in the model to estimate general cognitive ability (*g*), crystallized intelligence (cry), processing speed (spd) and memory (mem). Standardized factor loadings are displayed on the loading paths. Regression weights of latent factors onto $H_B$ are indicated as standardized estimates (*β*). **(b)** *β* coefficients estimated in SEMs using $H_B$, $H_{Se}$ and $D_{Se}$. The models fit the data well: $CFI > 0.95$, $SRMR < 0.08$ and $RMSEA < 0.08$ (see Supplementary Fig. 6). Here, '*' represents $p < 0.05$ and '**' indicates $p < 0.01$. **(c)** The distribution of individual static $H_B$. The blue lines ($H_B = -0.117$ and $0.123$) represent the cut-off values for a specific partition. The group sizes are also provided (red text). This SEM fits well with $CFI = 0.974$, $SRMR = 0.036$ and $RMSEA = 0.049$. **(d)** Estimated group-specific latent means for the four cognitive abilities. Cohen's *d* effect size estimates indicating group differences in cognitive abilities are displayed in red.

Importantly, the above reported group differences are robust for different partitions into three groups (Supplementary Fig. 7). Further partitioning the BG into two subgroups (i.e., resulting in four groups) also led to equivalent results (Supplementary Fig. 8). These findings provide robust evidence that higher general cognitive ability is associated with stronger integration, that higher crystallized intelligence and processing speed rely on stronger segregation, and that memory is the strongest in individuals at the balance between segregation and integration during rest.

**Discussion**

By proposing a hierarchical module approach to brain FC networks, we explicitly identified the functional balance between segregation and integration. Using the large-scale WU-Minn HCP dataset and a Gaussian linear model, we provided theoretical and empirical evidence that healthy young brains at rest are on average close to the balanced state. This state allows the brain to frequently switch between segregated and integrated configurations. Compared with graph-based network measures at a single level, our approach is more effective for revealing the intricate role of segregation, integration and their balance in different cognitive abilities across individuals. General cognitive ability is facilitated by higher global integration; better crystallized intelligence and processing speed are associated with higher segregation; and memory profits from the tendency towards the balance. Our results not only provide an effective analysis of hierarchical modules in brain FC networks, but also reveal the functioning principles of resting brains to support diverse cognitive demands by configuring the functional organization to segregation, integration or balance.

**Network segregation-integration balance of healthy young brains at rest.** The appealing hypothesis of resting brain at a balance between segregation and integration has been accepted by many researchers[3,4,12,16,17,26,38,39]. Although, several methods have been proposed to identify segregated and integrated brain states and the competition within and between modules is expected to capture the balance[9,17,40-42], developing a quantitative definition of the balance was still considered a great challenge. We argued that segregated and integrated brain activities are hierarchically organized across pronounced modules in FC networks[29,30], and found that eigenmodes can reflect the hierarchical modular partition. By characterizing the competition between hierarchical segregation and integration based on eigenmodes, we identified the explicit balance and provided theoretical and empirical evidence that brain functional organizations at rest are on average configured to a balanced state, but there are also significant individual variations in segregation, integration and balance. To our knowledge, this study provides the first quantitative evidence for the segregation-integration balance in large-scale resting brains of heathy young adults, although the hypothesis has been investigated for many years[3,4,12,16,17,26,38,39].

**Balance supports fast reconfiguration of brain's functional organizations.** Dynamical reconfiguration of brain functional organization between segregated and integrated states associates with diverse cognitive abilities and neurological disorders[12,14,41-43]. For example, Parkinson's disease is associated with a longer dwell time in segregated states and a lower number of transitions between segregated and integrated states[41-43], and individuals with higher intelligence dwells less often into states of particularly high network segregation[12]. However, previous studies used single-level methods (e.g., *k*-means clustering) to detect different brain states[14,41-43], which are unable to quantify a clear borderline for segregated and integrated states. Recently, Hilger et al. theoretically assumed a functional balance in a group of individuals ($M_{age} = 47.19$ years old) and regarded the group mean modularity as the borderline for identifying segregated and integrated states[12]. Beyond the theoretical assumptions, we provided a complete quantitative framework for identifying segregated and integrated states. We found that balanced brains work with a balanced time in segregated and integrated states and a highly frequent state switching. Thus, the functional balance allows for more flexible reconfiguration in brain functional organization[2], which is assumed to be necessary for the brain to transit from resting to task states in a timely manner. Our approach has great potential for future investigations of brain's dynamic reconfigurations and their relationships with cognitive abilities during aging, cognitive training and mental disorders.

**Network segregation, integration and their balance predict different cognitive abilities.** Our work demonstrated that segregation, integration and their balance predict different cognitive abilities. Even if general cognitive ability is supposed to be facilitated by the balance in NNT, system-wide evidence for such an association remains controversial[20-23]. For example, previous studies reported a relationship between general cognitive ability and global efficiency of resting FC networks[21,22], but a recent replication study with WU-Minn HCP data did not observe the relationship[23]. Our results further confirmed that the single-level network analysis cannot capture the relationship between network characterizations and general cognitive ability (Supplementary Table 1), but the hierarchical module analysis effectively revealed that general cognitive ability is robustly predicted by higher global integration. Since general cognitive ability in the present SEM is marked by reasoning tasks (fluid intelligence) (see Methods), our result is consistent with the NNT assumption that weak network connections between regions facilitate a "difficult-to-reach state", needed for fluid intelligence[25]. Thus, when individual's functional organizations are configured to integration, resting brains more strongly exhibit global cooperative activity and flexibly switches to a "difficult-to-reach state", supporting better fluid and general intelligence.

Crystallized intelligence is presumed to be facilitated by an "easy-to-reach state" in NNT[25]. Two studies with small sample sizes reported an association between global efficiency in FC networks and crystallized intelligence[21,22]. However, while using the NIH Toolbox Cognition Battery (NIHTB-CB) to assess crystallized intelligence, a recent replication study based on the WU-Minn HCP data found that crystallized intelligence was not associated with global efficiency, characteristic path length and global clustering coefficient[23]. Here, we used SEM to estimate crystallized intelligence as a latent variable, and observed a small but robust association between it and network properties. These associations were present for traditional single-level network measures and our multiple-level measures (Supplementary Table 1). Our results indicate a positive association between segregation and crystallized intelligence, providing further support for NNT[25]. Thus, an individual's tendency to exhibit more independent activity in specialized subsystems allows the network to function with an "easy-to-reach state" which predicts better crystallized intelligence.

Currently, NNT does not make any clear prediction for memory. Memory is itself a complex ability. The main dimensional distinction is being made between working, primary (short-term) and secondary (long-term) memory[25,44]. Indeed, specific memory task performances were shown in the literature to be facilitated by different segregated and integrated processes[45-47], such as the vivid memory requiring higher global integration than dim memory[47], implying that general memory may be facilitated by a balance between segregation and integration. Here we showed that memory is higher in individuals tending toward balance in resting brains. In cognitive and differential psychology, an influential perspective on working memory—taken to be the cognitive mechanism underlying general cognitive ability[48]—assumes it to be a system responsible for building relational representations through temporary bindings between mental chunks[49]. We here revoke to see general cognitive ability merely as a statistical summary of domain-specific cognitive abilities. And the general cognitive ability in our model was modelled as a factor marked by reasoning tasks (fluid intelligence) and a memory was nested under this general factor. The memory tasks in the WU-Minn

HCP arguably capture the ability of building, maintaining and updating arbitrary bindings[49,50], which we view as the basic cognitive mechanism underlying general cognitive ability. Thus, this ability is expected to be associated with a pronounced small-world topology[25], tending to display a balance between segregation and integration, as predicted in NNT[25]. Therefore, this study substantially validates and enriches the hitherto proposed NNT of human cognition.

Processing speed is another domain-specific ability that is not explicitly considered in NNT[25]. In terms of brain network characteristics facilitating processing speed, theoretically well justified predictions are difficult to be proposed[50,51]. Thus, results and their interpretation from previous studies mainly rely on motor performance. Here, we demonstrated that faster processing speed is associated with the tendency towards segregated activity in resting brains. This finding is consistent with theories aiming to understand the lifespan development of modularity[11]. More modular neural architectures are associated with better performance when short response deadlines are required in cognitive tasks of low difficulty. Thus, processing speed is expected to relate to stronger modularity, i.e., higher segregation. Furthermore, at the level of specific cognitive tasks, higher segregation was shown to be related to successful motor execution[7]. These theoretical claims and few empirical findings are consistent with our results showing that quicker processing speed is predicted by higher segregation. Our findings complement missing constituents of NNT that aim to provide a network neuroscience view on general and domain-specific cognitive abilities, to which processing speed arguably belongs.

**Outlooks.** This work also has several valuable outlooks. First, we demonstrate that resting brains are close to a balanced state, and that the functional balance is not simply beneficial for all cognitive abilities as sometimes assumed. Since resting brains were found to function around a critical state[18,52-54], the functional balance theoretically matches the criticality characterizing individuals on average. Thus, our results are in line with a recent finding in a local neural circuit that criticality is not optimal for easy tasks but can facilitate the processing of difficult tasks[55]. However, the relationship between the balance and criticality across individuals was not yet clearly understood, mainly due to lack of explicit identification of the functional balance. Our work thus provides a powerful tool to solve this pending issue in physics and network neuroscience.

Second, the functional balance is expected to provide potential for the brain to flexibly switch to task states so as to match the variable task demands. This hypothesis is consistent with emerging evidence that individual brains characterized by more efficient switching from resting to task states perform better on specific cognitive tasks[14,15,26]. We here found that the functional balance supports highly flexible transitions between segregated and integrated states, but this does not benefit all cognitive abilities, except for memory. Thus, complementing the present findings with additional task-state analysis in the future will help to explain whether the brain balance supports the efficient transition from resting to task states.

**Conclusion**

Altogether, we found that resting brains maintain a segregation-integration balance to support heterogeneous de-mands of diverse cognitive abilities, and across individuals, segregation, integration and their balance predict different cognitive abilities. The functional balance supports the best memory, higher segregation corresponds to better crystallized intelligence and processing speed, and higher integration is associated with better general cognitive abilities. This study not only contributes to testing current NNT claims, but also reframes this theory by including additional domain-specific abilities and a general cognitive ability factor with a straightforward psychological interpretation. Furthermore, the concepts proposed here are helpful to refine the methodology proposed for parameterizing the functional organization of the brain's dynamic activity, which has the potential utility in the rapidly growing field of network neuroscience focusing on aging, cognitive training and mental disorders.

**Methods**

**Dataset.** The magnetic resonance imaging (MRI) data consisting of 991 healthy young adults (female = 528, age range = 22~36 years) were extracted from WU-Minn HCP. This dataset contains structural MRI, diffusion MRI, resting state fMRI and behavioral measures on multiple cognitive tasks. Each participant completed a two-day measurement involving four high-resolution scanning sessions (time of repetition TR=0.72 s), with each session lasting for 864 s (1200 frames).

**Human brain connectomes.** Brain was parcellated into 360 regions according to the multimodal parcellation (MMP) atlas[56]. The blood-oxygen level dependent (BOLD) time series for each region was extracted with the standard procedure (see Supplementary Methods)[34]. The Pearson correlation coefficient between BOLD series of two regions was calculated to indicate the FC. For the stable average FC matrix, the BOLD series for four sessions were concatenated in all individuals so that we obtained stable FC across long enough time scales. For the individual static FC matrices, the BOLD series for four sessions in each individual were concatenated. For the dynamic FC matrices, the BOLD series for four sessions in each individual were first orderly concatenated, and then a sliding-time window method was applied. With a window width of 59.76 s (83 points) and a sliding step of 0.72 s (1 point), the concatenated long BOLD series was divided into 4717 small pieces to construct the temporal FC matrices. We set negative correlations to zero and applied no other operations to the FC matrices.

Probabilistic tractography was performed to extract the connection probability $p_{ij}$ from region $i$ to $j$ [56-58], and the SC between regions was computed as $w_{ij} = (p_{ij} + p_{ji})/2$. The SC matrix thus is:

$$A_{ij} = \begin{cases} w_{ij} & \text{for } i \neq j \\ 0 & \text{for } i = j \end{cases} \quad (1)$$

**Brain functional modes.** The FC matrix $C$ can be decomposed as $C = U \Lambda U^T$ with eigenvectors $U$ and eigenvalues $\Lambda$. In the spectral space, the eigenvalues $\Lambda$ are usually described as the contribution of functional modes to FC networks, and the total contribution $\sum_{i=1}^{N}\Lambda_i \equiv N$ is independent of the dynamical synchronizing process. However, as synchronization increases, cortical regions exchange more information, leading to stronger connectivity, accompanied by higher degree for the regions. In this case, the contribution of functional modes to FC networks needs to grow as well. Thus, we used $\Lambda^2$ to measure the contribution of functional modes to FC networks. Few eigenvalues had negative values and were set to zero.

**Hierarchical modular partition of FC networks.** Eigenmode-based analysis has been successfully applied to complex networks[32,59]. Here, we applied the NSP method to detect the hierarchical modules in FC networks. In the 1$^{st}$ mode, the elements in the eigenvector for all regions have the same sign, which was referred to as the first level with one functional module (i.e., whole-brain network). In the 2$^{nd}$ mode, the regions with positive signs in the eigenvectors were assigned as a module, and the remaining regions with negative signs were assigned as the second module, which was regarded as the second level distinguishing two functional modules. Each module in the second level can be further partitioned into two submodules based on the positive or negative sign of regions in the 3$^{rd}$ mode, constructing the third level. Successively, the FC network can be modularly partitioned into multiple levels with the order of functional modes increasing until a given level where each module involves a single region only. After each partitioning step, the regions were reordered, and the order within modules remained random. During this nested partitioning process, we obtained the module number $M_i(i=1,\cdots,N)$ and the modular size $m_j(j=1,\cdots,M_i)$ in each level.

**Hierarchical segregation and integration components.** Functional segregation and integration are intricately interrelated and hierarchically organized in a nested manner across multiple levels. This hierarchically segregated and integrated activity reflected by eigenmodes has the contribution $\Lambda^2$ to the functional organization (Fig. 1d). The first level in the FC network has only a single large module (Figs. 1b, c), reflecting global integration to allow effective communication across the whole brain and thus requires the largest contribution (Fig. 1d). The second level generates integration within the anterior or posterior module and segregation between them (Figs. 1a-c). This modular organization supports the strong communication and specialized processing within the anterior or posterior regions and weaker cooperation between them and thus requires less contribution compared to the first global mode (Figs. 1d). Consequently, higher-order modes with more modules and smaller modular sizes relate to deeper levels of finer segregated processes that generate more localized information flow and coordination, accompanied by lower contributions $\Lambda^2$ of the corresponding modes (Fig. 1d and Supplementary Fig. 1). Specifically, the levels with the highest module number (i.e., $M = N$), allowing independent activation of each region, indicate completely segregated activity and are associated with very small contributions (Fig. 1d and Supplementary Fig. 1). Thus, the functional modes with larger module numbers generate stronger segregation and smaller-scale local integration, producing weaker contributions to the functional organization.

Consistent with the graph-based modularity[14,27,33], modules at a given level support the segregation between them and integration within them. A larger module number $M_i$ reflects higher segregation at this level. Since this segregated and integrated activity makes the contribution of $\Lambda_i^2$ to the functional organization, the weighted module number in each level can be defined to reflect the hierarchical segregated and integrated interactions:

$$H_i = \frac{\Lambda_i^2 M_i}{N} \qquad (2)$$

where $N$ normalizes the module number $M$ to the range [0, 1]. At low-order levels, the module number $M_i$ is small and the contribution $\Lambda_i^2$ is large, corresponding to strong integration of smaller modules at higher-order levels into large modules at low-order levels. Meanwhile, the large modules are further integrated into even larger modules at lower-order levels, allowing us to quantify hierarchically nested segregation and integration. Thus, $H_i$ describes the nested segregation and integration across multiple levels.

However, the number of modules alone may not properly describe the picture of nested segregation and integration because the size of modules might be heterogeneous (Fig. 1d). Given an extreme case at the second level, for example, having two modules with a size comprising one region and $N$-1 regions, this level would produce very weak segregation and nearly global integration. The segregation becomes stronger if the modules have a more homogeneous size $m_j = N/M_i$. Thus, the segregation and integration component in each level needs to be corrected for heterogeneous modular sizes. The correction factor was calculated as $p_i = \sum_j |m_j - N/M_i|/N$, which reflects the deviation from the optimized modular size in the $i$th level. Then, $H_i$ was corrected as:

$$H_i = \frac{\Lambda_i^2 M_i (1-p_i)}{N} \tag{3}$$

This correction aims to reflect the influence of modular size. If modules are dominant with respect to their size at a given level, the integration within this level would become stronger and the segregation weaker, corresponding to smaller $H_i$. Should the deviation of modular size from homogeneity be large, the correction effect is stronger (see Fig. 1d).

At the first level, there is only a single module for the whole FC network, and this level was taken to calculate the global integration component:

$$H_{In} = \frac{H_1}{N} = \frac{\Lambda_1^2 M_1 (1-p_1)}{N^2} \tag{4}$$

Further normalization by the node number of $N$ results in a measure that is independent of the network size. Since the first level contains only one module, $p_1 = 0$, and the global integration component does not need to be corrected.

The total segregation component unfolds from the multiple segregated levels ($2^{nd}$-$N^{th}$ levels):

$$H_{Se} = \sum_{i=2}^{N} \frac{H_i}{N} = \sum_{i=2}^{N} \frac{\Lambda_i^2 M_i}{N^2}(1-p_i) \quad (5)$$

These definitions of the global integration and segregation components in the FC network are illustrated in Fig. 1d.

**Gaussian linear diffusion model.** In order to identify the theoretical balance of the brain, a Gaussian linear diffusion model was adopted. Let $x_i$ represent the neural activities of cortical regions that follow a Gaussian linear process[18,35,60]. The time evolution of neural population activities satisfies (see Supplementary Methods):

$$\frac{dx_i}{dt} = -x_i + c\sum_{j=1}^{N} A_{ij}(x_j - x_i) + \sqrt{2}\xi_i \quad (6)$$

where $c$ is the coupling strength between cortex regions, and $A$ is the brain SC matrix defined in Eq. (1). By averaging over the states produced by an ensemble of noise and defining the $Q = (1+cH)^{-1}$ where $H$ is the Laplace matrix of the SC matrix, the covariance of this model can be analytically estimated as[61,62]:

$$Cov = \langle XX^T \rangle = 2\langle Q\xi\xi^T Q^T \rangle = 2Q\langle \xi\xi^T \rangle Q^T = 2QQ^T \quad (7)$$

The simulated FC matrix $C$ can be calculated as:

$$C_{ij} = \frac{Cov_{ij}}{\sqrt{Cov_{ii}Cov_{jj}}} \quad (8)$$

Thus, built upon the Gaussian linear process of the fluctuating resting brain state, the simulated stable FC networks over long enough time can theoretically be obtained and compared with FC from real fMRI data by tuning the coupling parameter $c$.

**Calibration process.** We provided theoretical and numerical evidence for the resting brain to close to the balance between segregation and integration given a long enough fMRI time series. However, a previous study found that shorter fMRI series resulted apparently in higher segregation estimates in terms of larger modularity in FC networks[28] and we also observed stronger segregation in the case of shorter fMRI series lengths (see Supplementary Fig. 4). If this artifact were not taken into account, deviation measures from the dynamical balanced state would be biased towards more but artificial segregation. To address this limitation, the segregation and integration components in individual static FC networks need to be calibrated. Considering that the length of fMRI time series would mainly affect the segregation component in static FC networks (Supplementary Fig. 4), the group segregation component $H_{Se}$ and integration component $H_{In}$ were calibrated to the integration component $H_{In}^{S} = 0.18$ of the stable average FC network. This is equivalent to the mean integration component of simulated FC networks at the balanced state (i.e., $c = 70$) (Figs. 2g, h). The segregation components in brains with $H_B < 0$ are more sensitive to the fMRI length (Supplementary Fig. 4), and thus, a proportional calibration scheme was adopted. For individual static FC networks (obtained from four sessions), the vectors of segregation (or integration) components for 991 individuals are $H_{In} = [H_{In}^1, H_{In}^2, \cdots, H_{In}^{991}]$ and $H_{Se} = [H_{Se}^1, H_{Se}^2, \cdots, H_{Se}^{991}]$, and the calibrated results for each individual are $H_{Se}^{i'} = H_{Se}^{i} \frac{H_{In}^{S}}{\langle H_{Se} \rangle}$ and $H_{In}^{i'} = H_{In}^{i} \frac{H_{In}^{S}}{\langle H_{In} \rangle}$. Here, $\langle \ \rangle$ represents the group average across 991 individuals. After this calibration, the group average values of the segregation and integration components will be equal (i.e., $H_B = 0$), and the individual ranking of segregation and integration will be fixed.

For dynamic FC networks, the temporal segregation and integration components for each individual were respectively calibrated to its static segregation component $H_{Se}^{i'}$ and integration component $H_{In}^{i'}$ to maintain the individual ranking (Supplementary Fig. 4). The vectors of segregation (or integration) components for the $i$th individual across 4717 windows are $h_{Se}^{i} = [h_{Se}^1, h_{Se}^2, \cdots, h_{Se}^{4717}]$ and $h_{In}^i = \left[ h_{In}^1, h_{In}^2, \cdots, h_{In}^{4717} \right]$, and the calibrated results are $h_{Se}^{t'} = h_{Se}^t \frac{H_{Se}^{i'}}{\langle h_{Se}^i \rangle}$ and $h_{In}^{t'} = h_{In}^t \frac{H_{In}^{i'}}{\langle h_{In}^i \rangle}$. Here, $\langle \ \rangle$ represents the average across 4717 windows. This calibration maintains the individual ranking of static segregation and integration components, and the calibrated results are independent of the length of fMRI series (Supplementary Fig. 4). This individual calibration does not affect the dynamic results according to which the balanced brain is characterized by a balanced dwell time and the maximum transition frequency (Supplementary Fig. 9).

**Dynamic measures.** In order to characterize the dynamic properties in detail, the dwell time in segregated and integrated states was first defined as:

$$T_{In} = \frac{t_{H_B(t) \geq 0}}{t_{all}} \text{ and } T_{Se} = \frac{t_{H_B(t) < 0}}{t_{all}} \tag{9}$$

Here, $t_{H_B(t) \geq 0}$ and $t_{H_B(t) < 0}$ measure the duration of the dynamic process at $H_B(t) \geq 0$ and $H_B(t) < 0$, and $t_{all} = 3396.24\text{s}$ is the total time.

Second, the switching frequency was used to measure the transition speed between segregated and integrated states, which was defined as:

$$f_{IS} = \frac{n_{H_B(t)H_B(t+1) \leq 0}}{t_{all}} \tag{10}$$

where $n$ is the total number for time $t$ satisfying $H_B(t) H_B(t+1) \leq 0$.

Third, the competition between dynamic segregated and integrated states across long time periods may be balanced, but the fluctuations can vary much at temporal windows. The amplitude of dynamic deviation from the balanced state to the integrated state or segregated state was thus calculated as:

$$D_{In} = \frac{\sum H_B(t)|_{\geq 0}}{t_{H_B \geq 0}} \text{ and } D_{Se} = \left| \frac{\sum H_B(t)|_{<0}}{t_{H_B < 0}} \right| \tag{11}$$

Here, $H_B(t)|_{\geq 0}$ and $H_B(t)|_{<0}$ represent the positive and negative $H_B$ during the dynamic process. A smaller $D_{In}$ (or larger $D_{Se}$) indicates that the brain deviates towards a segregated state with a higher amplitude.

**Structural Equation Modeling (SEM).** Cognitive behavioral measures were collected from nine specific tasks (see Supplementary Table 2): Picture Sequence Memory (PicSeq), Penn Word Memory Test (IWRD), Penn Progressive Matrices (PMAT), Variable Short Penn Line Orientation Test (VSPLOT), Picture Vocabulary (PicVocab), Oral Reading Recognition (ReadEng), Dimensional Change Card Sort (CardSort), Flanker Task (Flanker) and Pattern Completion Processing Speed (ProcSpeed). To obtain estimates of general and domain-specific cognitive abilities, we applied a bi-factor SEM minus one to extract latent factors of common phenotypes that explain variability across the above listed tasks[63,64] (see Supplementary Methods). According to Ref. [[63]], we modeled crystallized intelligence by including ReadEng and PicVocab as indicators, memory ability based on PicSeq and IWRD and processing speed based on CardSort, Flanker and ProcSpeed. Especially, general cognitive ability was modeled as the shared variance across a broad set of task performance scores. But to avoid anomalous estimation results associated with bi-factor models[64], PMAT and VSPLOT were only loaded onto general cognitive ability which can then be interpreted as reasoning (fluid intelligence), required by all tasks to a different degree. This interpretation is consistent with the well-established finding that general cognitive ability and fluid intelligence are correlated above .93[48].

According to the above model structure (see Fig. 5a), we further conducted a multigroup SEM to investigate nonlinear associations. We compared the latent ability in different groups according to the magnitude of segregation, balance, or integration (Fig. 5c). We first partitioned 991 individuals into SG, BG and IG according to a set of thresholds in sorted individuals according to their $H_B$. In Fig. 3c, we orderly selected the individuals starting from the largest $H_B$. The group size of IG was first selected with $N_{IG}$ individuals. Then, starting with the $(N_{IG}+1)$th person, we selected the individuals for the BG. By constraining the average $H_B = 0$ in the BG, we can fix the group size of IG with $N_{BG}$. Individuals at the left were assigned to the SG. After the partition, we estimated group-specific model parameters using multigroup SEM. Different partitions were explored by a scanning step of 10 persons in the IG to assess the robustness of results across different partitioning thresholds (see Supplementary Figs. 7, 8).

The SEM analysis was performed with the lavaan package in R[65]. The comparative fit index (CFI), root mean square error of approximation (RMSEA) and standardized root mean-square residual (SRMR) were used to evaluate the fit of the estimated models[57,66]. Specifically, the model was considered to fit well for CFI>0.95, SRMR<0.08 and RMSEA<0.08[66]. In multigroup SEMs, we compared the goodness-of-fit of the model with freely estimated average latent ability with a model introducing equality constraints on the latent means across the groups. The $\chi^2$ difference test was used for model comparison. The resulting $p$-value of the test statistic indicates whether the unconstrained model was a significantly better fit than the constrained model. Group comparisons were based on Cohen's $d$ ($d \geq 0.2$ reflects a small effect size, $d \geq 0.5$ represents a medium effect size and $d \geq 0.8$ indicates a large effect size).

**Graph-based network measures.** To compare the results between graph-based network measures and hierarchical measures of functional segregation and integration in FC networks, the modularity and participation coefficient were computed as well. The modularity $Q$ in undirected weighted networks is[33]:

$$Q = \frac{1}{l} \sum_{i,j \in N} \left[ w_{ij} - \frac{k_i k_j}{l} \right] \delta_{m_i, m_j} \qquad (12)$$

Here, $w_{ij}$ is the connectivity between nodes $i$ and $j$, $k_i = \sum w_{ij}$ is the degree of node $i$ and $l$ is the sum of all weights in the network. $\delta_{m_i,m_j}=1$ if nodes $i$ and $j$ are in the same module; otherwise, $\delta_{m_i,m_j}=0$. The modularity $Q$ quantifies the degree to which a network is decomposed into densely connected modules, and a larger $Q$ reflects higher segregation.

The participation coefficient quantifies the degree to which a node is connected to other nodes across diverse modules. Its definition for a node is[33]:

$$P_i = 1 - \sum_{m=1}^{M}\left(\frac{k_i(m)}{k_i}\right)^2 \quad (13)$$

where $M$ is the module number and $k_i(m)$ is the connectivity strength of node $i$ within module $m$. The participation coefficient $P$ of a network is the average of $P_i$ in all nodes. Here, the modules were previously defined according to the seven functional subsystems[37]. These network measures were calculated with the Brain Connectivity Toolbox (BCT, https://www.nitrc.org/projects/bct).

**Data availability**

The datasets that support the findings of this study are available at http://www.humanconnectome.org/study/hcp-young-adult.

**Code availability**

The codes used in this study are available at https://github.com/TobousRong/Hierarchical-module-analysis.

**Acknowledgments**

This work was supported by the National Natural Science Foundation of China (Nos. 11802229, 11972275, 11772242), the Hong Kong Scholars Program (No. XJ2020007), the Outstanding Youth Science Fund of Xi'an University of Science and Technology (No. 2019YQ3-11), the Hong Kong Baptist University Research Committee Interdisciplinary Research Matching Scheme 2018/19 (IRMS/18-19/SCI01) and Germany-Hong Kong Joint Research Scheme (G_HKBU201/17 awarded to C.Z. and ID 57391438 awarded to A.H.). This research was conducted using the resources of the High Performance Computing Cluster Centre, Hong Kong Baptist University, which receives funding from RGC, University Grant Committee of the HKSAR and HKBU.

**Author contributions**

C. Z., A. H and R. W. conceived the study; A. H and M. L. built the SEM; M. L. processed the MRI data and SEM; R. W. programmed numerical simulations of Gaussian linear diffusion model; C. Z., A. H., R. W. and M. L. wrote the manuscript; R. W and X. C. prepared the figures and manuscript format, and all authors edited the manuscript.

**Competing interests**

The authors declare no competing interests.